\documentclass[usenatbib]{mn2e}

\usepackage[dvips]{graphicx}
\usepackage{amssymb}
\usepackage{txfonts}
\usepackage{colordvi}
\usepackage{url}

\def \pwr {{\rm ergs\,s^{-1}}}

\newbox\grsign \setbox\grsign=\hbox{$>$} \newdimen\grdimen \grdimen=\ht\grsign
\newbox\simlessbox \newbox\simgreatbox \newbox\simpropbox
\setbox\simgreatbox=\hbox{\raise.5ex\hbox{$>$}\llap
     {\lower.5ex\hbox{$\sim$}}}\ht1=\grdimen\dp1=0pt
\setbox\simlessbox=\hbox{\raise.5ex\hbox{$<$}\llap
     {\lower.5ex\hbox{$\sim$}}}\ht2=\grdimen\dp2=0pt
\setbox\simpropbox=\hbox{\raise.5ex\hbox{$\propto$}\llap
     {\lower.5ex\hbox{$\sim$}}}\ht2=\grdimen\dp2=0pt

\title[Bow-shock PWNe]{Bow-shock Pulsar Wind Nebulae Passing Through Density Discontinuities}

\author[D. Yoon et al.]
{Doosoo Yoon$^{1,2}$ and Sebastian Heinz$^2$\\
$^1$Shanghai Astronomical Observatory, Chinese Academic of Science, China\\
$^2$Department of Astronomy, University of Wisconsin-Madison, Madison, WI, USA
}

\date{Accepted 2016 October 6; in original form 2016 June 17}

\pagerange{\pageref{firstpage}--\pageref{lastpage}}
\pubyear{2016}

\begin{document}

\maketitle

\label{firstpage}

\begin{abstract} 
  Bow-shock pulsar wind nebulae are a subset of pulsar wind nebulae
  that form when the pulsar has high velocity due to the natal kick
  during the supernova explosion. The interaction between the
  relativistic wind from the fast-moving pulsar and the interstellar
  medium produces a bow-shock and a trail, which are detectable in
  H$_{\alpha}$ emission.  Among such bow-shock pulsar wind nebulae,
  the Guitar Nebula stands out for its peculiar morphology, which
  consists of a prominent bow-shock head and a series of bubbles
  further behind. We present a scenario in which multiple bubbles can
  be produced when the pulsar encounters a series of density
  discontinuities in the ISM. We tested the scenario using 2-D and 3-D
  hydrodynamic simulations.  The shape of the guitar nebula can be
  reproduced if the pulsar traversed a region of declining low
  density.  We also show that if a pulsar encounters an inclined
  density discontinuity, it produces an asymmetric bow-shock head,
  consistent with observations of the bow-shock of the millisecond
  pulsar J2124-3358.
\end{abstract}


\section{Introduction}

Pulsar Wind Nebulae (PWNe) are produced when a pulsar's relativistic wind interacts with the
surrounding medium. They produce emission across a broad spectral range from radio synchrotron
emission to $\gamma$-rays [\citet{Gaensler:06} and references therein]. A subset of PWNe has been
identified in which pulsars with high spatial velocities produce bow-shock structures and cometary
shapes [e.g., B1957+20 \citep{Kulkarni:88}; J0437-4715 \citep{Bell:95}; RXJ1856.5-3754
\citep{Kerkwijk:01}; B0740-28 \citep{Jones:02}; J2124-3358 \citep{Gaensler:02}; J1747-2958 (the
``mouse'' nebula) \citep{Gaensler:04}]. Such bow-shocks are detectable in collisionally excited
H$_{\alpha}$ emission \citep{Kerkwijk:01,Jones:02}. A theoretical model of Balmer-dominated pulsar
bow-shocks was developed by \citet{Kulkarni:88}.

A pulsar's high-spatial velocity is caused by the natal kick of the neutron star during its
supernova explosion. Numerous models have been proposed to take into account the deviation from
spherical symmetry at the core of the progenitor star. This deviation can occur because of
anisotropic neutrino emission during the Kelvin-Helmholtz cooling of the nascent remnant
\citep{Sagert:08}, anisotropic mass ejection during the SN explosion due to standing-accretion shock
instability \citep{Blondin:03,Hanke:12}, or convective shell-burning during the final life cycle
stage of the progenitor star \citep{Burrows:96}.

\subsection{Young bow-shock nebulae}

\citet{Bernstein:09} carried out special relativistic, hydrodynamics simulation of pulsar wind
bow-shock nebulae, showing that the relativistic backflow from the fast-moving pulsar thermalizes in
the form of a slowly inflating, trailing spherical bubble. This energy-driven bubble model also
applies to bow-shock nebulae driven by jets from low-mass X-ray Binaries moving through the
interstellar medium (ISM) \citep{Yoon:11}.

\begin{figure}
    \centering
    \includegraphics[width=\columnwidth]{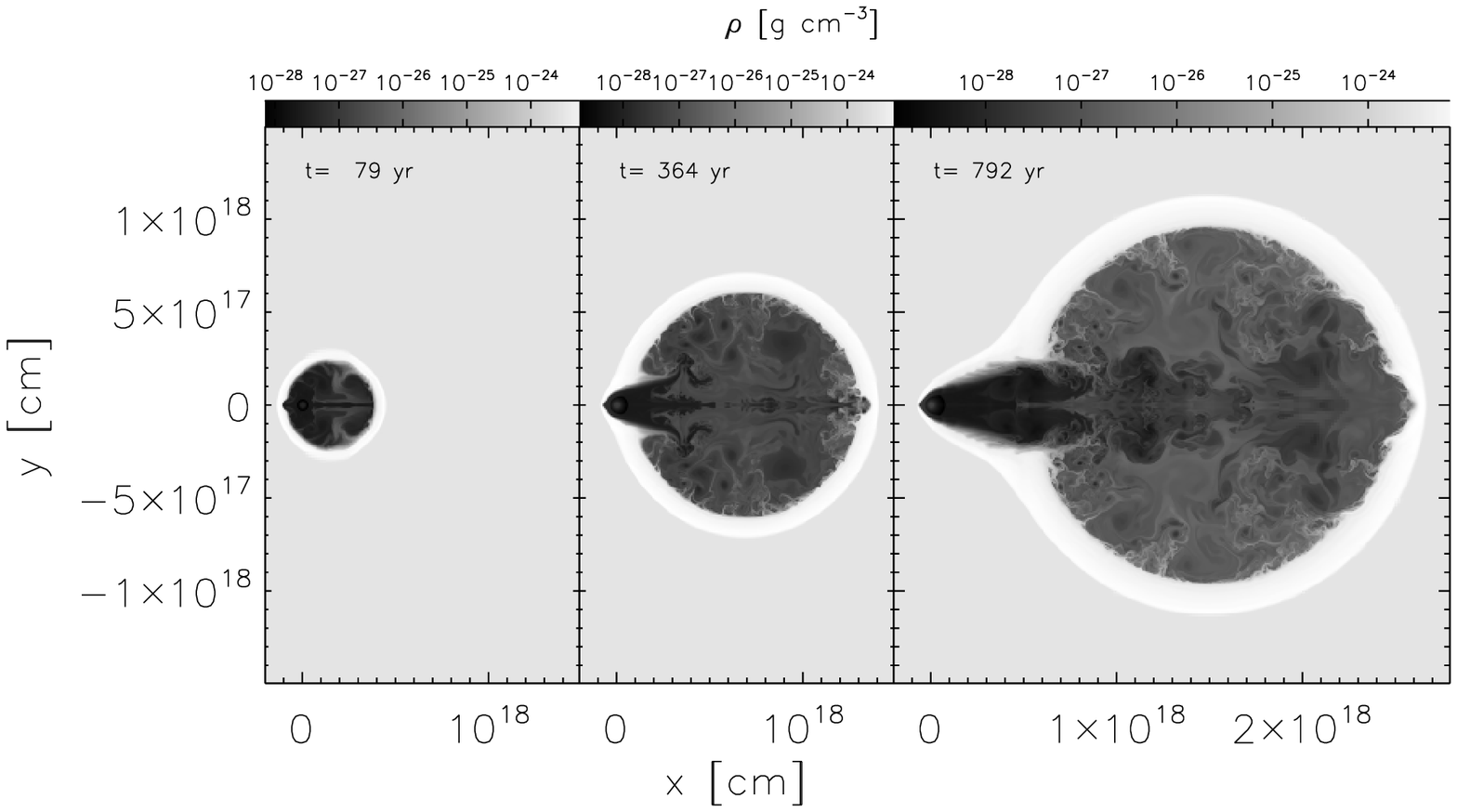} \\
    \includegraphics[width=\columnwidth]{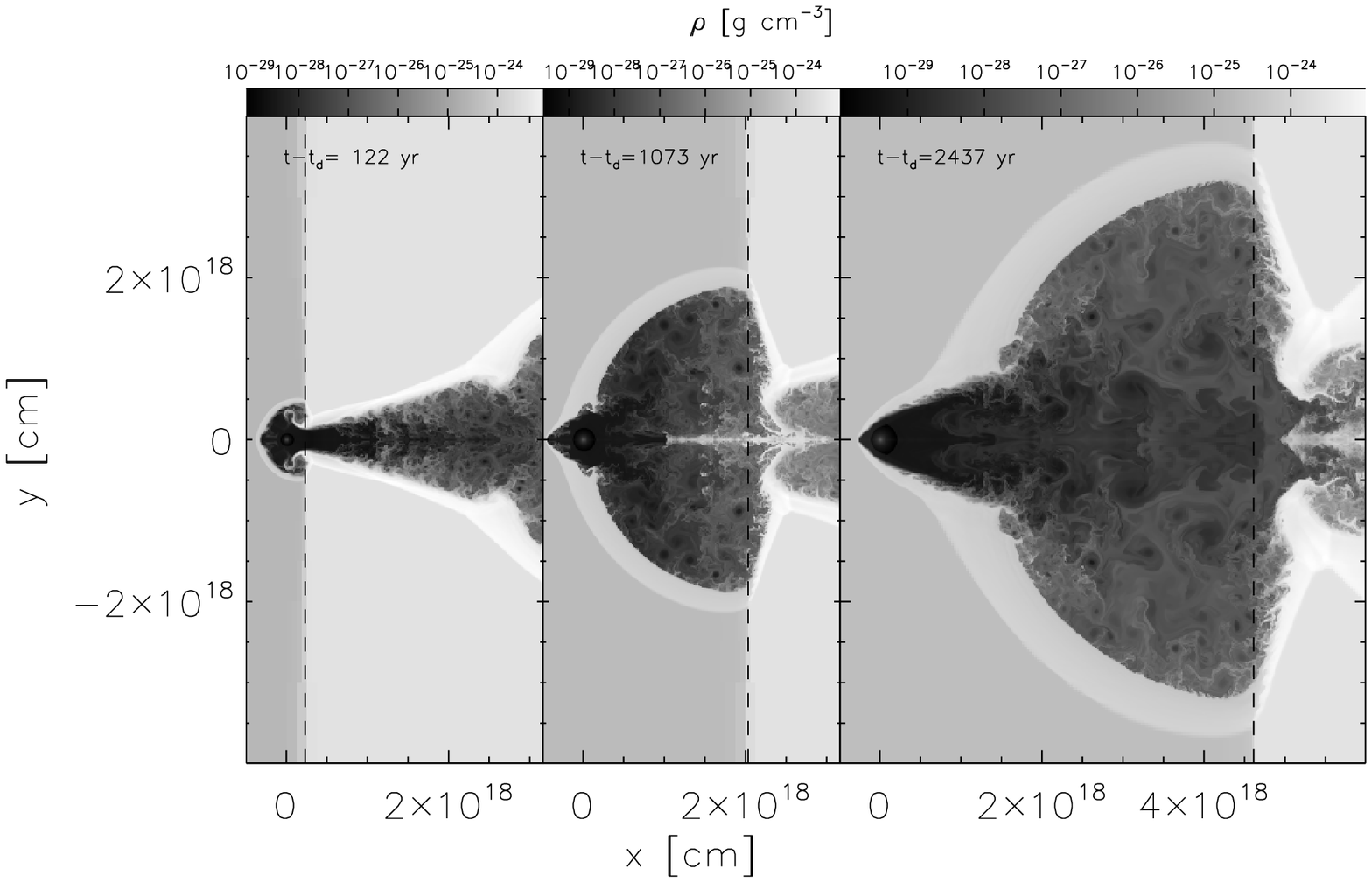} 
    \caption{Density slice showing the evolution of
      expanding pulsar-wind driven bubbles
      caused by the initial explosion (upper panel)
      and the passage of a density discontinuity
      (lower panel). The vertical dashed lines in the lower case
      indicate the location of the discontinuity.  The ambient
      densities are $\rho_{0}=1.67\times10^{-24}\,{\rm g\,cm^{-3}}$
      (upper case), and
      $\rho_{0}=1.67\times10^{-24}\rightarrow 1.67\times10^{-25}
      \,{\rm g\,cm^{-3}}$
      (lower case). For both cases, the pulsar is still inside the
      bubble in left-most plot, breaks out of the
      bubble in middle plot, and interacts with the
      ISM producing bow-shock in right most plot.}
\label{fig:evol}
\end{figure}

This model describes the evolution of bow-shock PWNe during the earliest stage, when the pulsar's
location is close to its birth-place.  The formation of the bow-shock/bubble morphology of bow-shock 
nebulae is illustrated in the top panel of Figure~\ref{fig:evol} and can be understood
as follows: When a pulsar wind turns on, it inflates an energy-driven bubble, following the
well-known solution by \citet{Castor:75}, the continuously fed analog to the Sedov-Taylor
solution. From simple dimensional arguments it is easy to see that the radius of such a bubble
is described by the relation
\begin{equation}
  R_{\rm bubble} \sim \left(\frac{\dot{E}\,t^{3}}{\rho_{\rm
        ISM}}\right)^{1/5}
  \label{eq:bubble2}
\end{equation}
where $\dot{E}$ is a spin-down loss energy rate, and $\rho_{\rm ISM}$ is an ambient density.
The expansion velocity of a wind-driven bubble decreases with time as $v_{\rm b} \propto t^{-2/5}$.
Initially, the expansion velocity of this bubble must therefore be faster than the kick velocity
of the pulsar.  However, the velocity will eventually drop below the pulsar's velocity, and
the pulsar will break out of the bubble.  Once this is the case, the pulsar wind will form a
classical bow-shock nebula, where, on the up-wind side, the ram pressure of the ISM balances
the wind pressure.

As described in \citet{Bernstein:09} and \citet{Yoon:11}, the bow-shock will have a roughly
conical shape behind the pulsar and matter will be accelerated from the high-pressure bow-shock
back in the direction of the bubble. The bubble will therefore continue to be inflated while
the pulsar moves away from its center.

For young pulsars, one might therefore expect the presence of such a neck/bubble structure in
H-alpha images.  However, as we will argue, the presence of bubbles behind the pulsar in later
stages of pulsar evolution can not be explained by a model in which the bubble is inflated by the
initial plerion after the pulsar turned on, given the large distances older pulsars will have
traveled through the Galaxy since turning on.  These bubbles or cavities are not rare [e.g. PSR
J2030+4415, PSR J1509+5850 \citep{Brownsberger:14}, PSR B2224+65 \citep{Chatterjee:02}].
Observations of bubble- and shell-like structures behind bow-shock PWNe in older pulsars must
therefore be explained by a different mechanism.

\subsection{The Guitar Nebula}

The Guitar Nebula is one of the most spectacular PWNe, produced by one of the fastest known
pulsars, PSR B2224+65, which has a transverse velocity of $v_\perp \ge 1000 \, \rm km\,s^{-1}$.
In H$_{\alpha}$ observations, the nebula has a guitar-like shape with a bright head, trailing
neck, and a series of roughly circular shells/bubbles \citep{Cordes:93, Chatterjee:02}, shown
in Figure~\ref{fig:guitarHa}.

The pulsar most probably originated from the Cygnus OB3 association about 0.8 Myr ago
\citep{Tetzlaff:09}, thus the observed bubbles behind it are clearly {\em not} generated from
the pulsar's birth.

From eq.~(\ref{eq:bubble2}), a rapid change in $\dot{R}_{\rm bubble}$ can be brought about by
two things:
\begin{itemize}
\item{A significant, instantaneous increase in the pulsar wind power
    will generate a new spherical bubble at the location where the
    wind power increased.}
\item{A sharp decrease in the ISM density $\rho$ will also lead to the
    inflation of a new bubble.}
\end{itemize}

Time-variable $\dot{E}$ fails to account for the brightening in the constricted region of the
Guitar Nebula, and the amount of the variability in $\dot{E}$ is likely negligible compared
to large morphological changes of the nebula \citep{Chatterjee:04}.  Thus, we conclude that
variation of the ambient density is likely responsible for the curious morphology of the
Guitar Nebula\footnote{It had been suggested that dynamical instabilities in bow-shock backflow
could produce similar results \citep{Kerkwijk:08}. As we will show below, our 3-D simulations
should capture such instabilities and show no sign of the kind of structure observed in the
guitar nebula in the absence of a density discontinuity.}.  As an another possible mechanism,
\citet{Morlino:15} suggested that mass-loading of encountered neutral hydrogen makes bow-shock
unstable likely altering the shock structures and producing the bubbles.

The well-documented multi-phase nature of the ISM supports the notion that bow-shock
pulsars will sample a range of densities along their trajectories. For example, small-scale
density variations corresponding to tens of AU are observed in the Galactic HI absorption
\citep{Deshpande:00,Faison:01}. \citet{Patat:10} also detect interstellar medium column
density variations on scales of $\sim 100$ AU, inferred from expansion rates of a Type Ia
supernova photospheres.

The question is: Can the morphological details of the Guitar Nebula be explained by a reasonable
density structure of the ISM the pulsar is traversing?

The primary goal of this study is therefore to investigate the dynamic evolution of bow-shock
PWNe in inhomogeneous environments. \citet{Noutsos:13} reported that the spin-down ages of
$\tau_{c} = P/(2\dot{P})$ and kinematic ages $t_{\rm kyn}$ of a sample of pulsars are within
the range $10^5 \sim 10^7\,\rm yr$.  Under the reasonable assumption that pulsars are born
somewhere within $100 \,\rm pc$ of the galactic mid-plane, they are expected to intersect the
galactic plane at least once in the time travel. It is quite possible that fast-moving pulsars
encounter significant changes in their environment.

In this work, using hydrodynamic simulations, we test whether a bubble forms when a pulsar
undergoes a sharp change in the ambient density, and examine the temporal evolution of the bubble
and the bow-shock after such a change.  We also develop an analytic formula for the bow-shock to
discuss the morphological variations of the structure of the shock in response to density changes
in order to provide observational diagnostics for observations of bow-shock PWNe.

We apply this model to the Guitar Nebula, suggesting possible dynamical evolutionary paths.  The
organization of this paper is as follows. In Section 2, we discuss the numerical method and the
initial setup. In Section 3, we discuss the evolution of the PWNe encountering density
discontinuities. In Section 4, we compare our numerical results with H$_{\alpha}$ observations of
the Guitar Nebula.  We also discuss the effects of an inclined density transition layer to explain
the asymmetric bow-shock in PSR J2124-3358.  In Section 5, we summarize our results.

\section{Numerical method}

\subsection{The code}

Simulations were carried out with the FLASH 3.3 hydrodynamic code \citep{Fryxell:00}, which is a
modular and parallel simulation code suitable for solving compressible flow problems. An adaptive
mesh refinement (AMR) algorithm in the code enables us to investigate the global evolution of
PWNe efficiently using moderate computing resources. 
The refinement level of the simulations is determined by both the grid position and physical
quantity: We set the finest resolution at the pulsar to safely resolve the pulsar wind and the
bow-shock structure, which is \emph{lrefine\_max}=11 implying that the highest resolution is
$1.2\times 10^{14}\,{\rm cm}$.

We adopt the piecewise-parabolic solver
for our non-relativistic hydro runs \citep{Colella:84}. We use a multi-gamma equation of state,
where the adiabatic index of the ambient medium is $\gamma_{\rm ambient} = 5/3$ and that of the
pulsar wind is $\gamma_{\rm pulsar} = 4/3$.  We verified that the (non-relativistic) bow-shock
and bubble are well described by our non-relativistic simulations, even though the pulsar wind
itself is relativistic, by performing a test study with fully relativistic simulations.

\subsection{Implementation of the Pulsar Wind}\label{subsec:pulsarwind}

The bulk of a pulsar's spin-down loss energy is converted into a relativistic pulsar wind
\citep{Michel:69}.  In general, the wind is likely anisotropic, and the anisotropy of the wind
momentum flux should be taken into account to explain detailed bow-shock structures
\citep{Vigelius:07}. However, the global features produced by interactions between pulsar winds and
the surrounding medium are less influenced by the anisotropy of the wind.  Moreover, there is no
evidence for an anisotropic or clumped pulsar wind in the series of H$_{\alpha}$ observations of the
Guitar Nebula \citep{Chatterjee:04}.  Therefore, we approximate the pulsar wind as a spherical
inflow, injecting flux isotropically through a spherical surface at the pulsar's location for
numerical simplicity (see Section 4).

A pulsar wind is ultra-relativistic, interacting with the ambient medium of either the supernova
remnant (SNR) or the ISM, and the Lorentz factor is in the range of $10^{4}\sim10^{7}$
\citep{Kennel:84}. The pulsar wind produces an expanding bubble of relativistic particles around the
neutron star.  In principle, to model the {\em internal} structure and properties of the wind,
relativistic and/or magnetohydrodynamic (MHD) simulations are necessary.  For instance,
\citet{Komissarov:04,Zanna:04} studied the evolution of PWNe by using relativistic MHD simulations,
and reproduced the polar jet from the pulsar by magnetic hoop stress, which is observed in the X-ray
images of Crab Nebula \citep{Weisskopf:00}. \citet{Bucciantini:05} presented the effect of wind
magnetization on the evolution of shock layers in bow-shock PWNe, and constraint the flow velocities
in the down-wind streams.  \citet{Bernstein:09} focused on studying the inflating bubble behind
pulsars by using relativistic hydrodynamic simulations. They concluded that the shockwaves provide
the thermalized energy into the bubble, allowing it to expand continuously.  However, because the
evolution of large-scale bubbles and bow-shock are non-relativistic, we carried out purely
non-relativistic hydrodynamic simulations, for reasons of computational feasibility.  We compared
our results with relativistic pulsar wind model in \citet{Bernstein:09}, and it shows no noticeable
differences in the dynamics of PWNe.  To verify that the non-relativistic evolution of the bow-shock
and bubble are correctly reproduced by our simulations, we carried out a test run with a fully
relativistic algorithm that will be discussed in \S \ref{subsec:relativistic}.

When a pulsar that loses energy at the rate of $\dot{E}$ moves through an ambient medium with
density of $\rho_{0}$ at velocity $v_{\star}$, the wind will generate a bow-shock with the standoff
distance $R_{0}$, at the stagnation point where the pulsar wind momentum flux is balanced by
the ISM's ram pressure:
\begin{equation}\label{eq:std}
  R_{0} = \left( \frac{\dot{E}}{4 \pi \rho_{0} v_{\star}^{2} v_{\rm wind}  } \right)^{1/2},
\end{equation}
where the pulsar wind velocity in our case is
$v_{\rm wind} = 10^{10} \, \rm cm \, s^{-1}$.  

The pulsar wind is injected isotropically at a spherical boundary of radius $R_{\rm nozzle}$
centered on the coordinate origin in our grid.  We refer to this injection region as the
{\em nozzle}.  Because we inject the wind at asymptotic velocity, the injection radius $R_{\rm
nozzle}$ of the nozzle is arbitrary, but must be smaller than the standoff distance $R_{0}$,
which we verified for each simulation.

Values of $\dot{E}$ for the observed pulsar population are in range of $10^{28} \sim 10^{38} \,\pwr$
\citep{Manchester:05}, and pulsars with $\dot{E} \gtrsim 10^{36}\,\pwr$ produce prominent PWNe
\citep{Gotthelf:04}.  To simulate typical bow-shock PWNe, we set $\dot{E}=10^{36}\,\pwr$ for all
general cases investigated, except for the Guitar nebula model which has a spin-down loss energy
rate, $\dot{E}=10^{33}\,\pwr$ \citep{Chatterjee:04}. To resolve the pulsar wind and bow-shock
structure appropriately, we set the (adaptive) numerical resolution around the pulsar to have at
least 10 cells across $R_{\rm nozzle}$.

\subsection{Initial Setup and Boundary Conditions}

Rather than simulating a moving pulsar in a stationary medium, we fix the pulsar's location
at $x=0$ for all runs. The external medium is streaming past the pulsar with a fixed velocity
$v_{\rm x}$ of 300, 600, or 900 $\rm km\,s^{-1}$ (see Table~\ref{tab:param}).  For the Guitar
nebula model, we set the velocity to 1,500 $\rm km\,s^{-1}$.  We apply inflow boundary conditions
with velocity $v_{\rm x}$ and ISM density and pressure at the lower $x$-boundary. These velocities
represent the actual proper speeds of the pulsars.

Following \citet{Cox:05}, the pressure and density in the ambient medium are set to typical
values in the galactic disk, $P_{0}=3 \times 10^{-12} \rm \,erg\,cm^{-3}$ and $n_{\rm ISM}=1\rm
\,cm ^{-3}$ for most cases. The ISM temperature is between $10^{4}\sim 10^{5}$ K.

As discussed above, the pulsar produces a spherical bubble in the early stage of the simulation.
After a break time, $t_{\rm break}$, the pulsar breaks out of the bubble producing a bow-shock
structure.  Because the goal of this study is to understand bow-shock evolution when a pulsar
propagates through a non-uniform density medium, the density gradient/discontinuity is introduced at
the left boundary with varying boundary condition after the bow-shock structure is well
developed,$t>t_{\rm break}$, and converged into the analytically expected shape
[eq.~(\ref{eq:dwdx})]; and it is correctly advected downstream towards the bow-shock.

We examined four configurations of the density change: sharp or secular changes, and either a
ten-fold increase or a ten-fold decrease.  For all cases, we adopt the same value of distance,
$d=4.12\times10^{18}\rm\,cm$, between the location of the density discontinuity and initial
location for the purpose of comparison. The detailed information we use for this study is listed
in Table~\ref{tab:param}.

It is important to note that the density configurations studied in this work are
over-simplifications: The interfaces of real density jumps will not be plane, the transitions may be
more or less gradual, and the media on either side of the transition may not be uniform. The goal of
this study is to investigate the general behavior of pulsar wind bow shock nebulae in multi-phase
gases and to explore the kinds of shapes the shocks can take, and simple density configurations are
best suited for providing a robust, simple view of this problem.

\setlength{\tabcolsep}{0.1in}
\begin{table*}
  \centering
  \caption{Parameter of the Simulations}
  \vspace{0.1cm}
  \label{tab:param}
    \begin{tabular}{cccccccc} \hline \hline
        Model & $v_{\star}\,[\rm km\,s^{-1}]$ & $\dot{E}_{36}^{a}$ & density ratio$^{b}$ & $R_{\rm nozzle}\,[\rm cm]$ & transition width ($l_{\rm tr}$) & dimension & incident angle \\ \hline
        L10\_300     & 300  & 1 & 0.1 & $1.2\times10^{16}$ & instant & 2-D & 90$^{\circ}$ \\
        H10\_300     & 300  & 1 &10 & $1.2\times10^{16}$  & instant & 2-D & 90$^{\circ}$ \\ 
        L10\_600     & 600  & 1 &0.1 & $1.2\times10^{16}$  & instant & 2-D & 90$^{\circ}$ \\
        L10br\_600   & 600  & 1 &0.1 & $1.2\times10^{16}$  & $5\times10^{17}$ cm & 2-D & 90$^{\circ}$ \\
        H10br\_600   & 600  & 1 & 10 & $1.2\times10^{16}$  & $5\times10^{17}$ cm & 2-D & 90$^{\circ}$ \\
        Uniform\_600 & 600  & 1 & 1 & $1.2\times10^{16}$  & - & 2-D & - \\
        L10\_900     & 900  & 1 &0.1 & $1.2\times10^{16}$  & instant & 2-D & 90$^{\circ}$ \\
        H10\_900     & 900  & 1 &10  & $1.2\times10^{16}$  & instant & 2-D & 90$^{\circ}$ \\
        Guitar\_LOS90       & 1500 & 0.001 & three changes$^{c}$ & $5\times10^{15}$ & mixed$^{d}$ & 2-D & 90$^{\circ}$ \\
        Guitar\_LOS60       & 1730 & 0.001 & three changes$^{c}$ & $5\times10^{15}$ & mixed$^{d}$ & 2-D & 90$^{\circ}$ \\
        L10\_1000\_5deg     & 1000 & 1 & 0.1 & $5\times10^{15}$ & $4.4\times10^{16}$ cm & 3-D & 5$^{\circ}$ \\
        L10\_1000\_45deg    & 1000 & 1 & 0.1 & $5\times10^{15}$ & instant & 3-D & 45$^{\circ}$ \\ \hline
    \end{tabular}
    \begin{itemize}
        \item[a] $\dot{E}_{36}= \dot{E}/(10^{36}\,\pwr)$ 
        \item[b] Ratio of the final ambient density to the initial density.
        \item[c,d] There are three changes in the ambient density for
          our fiducial Guitar nebular model. The ambient density
          changes to $\rho_{0,{\rm init}}/5$, further decreases to
          $\rho_{0,{\rm init}}/10$, and then finally increases to
          $\rho_{0,{\rm init}}$. The first and third changes occur
          instantly, and the second change occurs smoothly with the
          transitional length of
          $1.3\times10^{17} / \sin{\theta_{\rm LOS}} \,{\rm cm}$,
          where $\theta_{\rm LOS}$ is the viewing angle between the
          pulsar's velocity vector and the line of sight.
    \end{itemize}
\end{table*}

Simulations were primarily carried out in axi-symmetry, i.e., 2-D, for the purpose of saving
computing resources.  We carried out two types of 3-D simulations: one for validating 2-D runs,
and a set of 3-D runs for studying the non-axisymmetric case (see Table~\ref{tab:param}). The
latter case investigates a density discontinuity at an angle relative to the pulsar velocity.
We investigated two cases: inclination angles of $45^{\circ}$ and $5^{\circ}$, respectively,
and the result will be discussed at \S\ref{subsec:asym}.

\section{Results}

\subsection{Bow-shock Head}

The global evolution of the down-wind shock structure is generally similar to that of supersonic
low-mass X-ray binaries (LMXBs) described in \citet{Yoon:11}, despite the different engine in the
two systems. Around the pulsar, the unperturbed pulsar wind propagates to the termination shock
where the wind slows down to subsonic speed because of its interaction with the ambient medium.
The bow-shock becomes quasi-conical on the down-stream side (with an opening angle comparable
to the Mach angle).

Within the down-wind cone, the shocked pulsar wind propagates away from the pulsar due to the
pressure gradient behind the bow-shock.  The tail continues to drive the conical down-wind bow-shock 
as long as its pressure exceeds the thermal pressure in the ambient medium. The bow-shock
tail can be described as a narrow (cylindrical) cavity filled with relativistic exhaust from
the pulsar.

As discussed in \citet{Yoon:11}, the shape of the cavity along the $x$-axis can be described
analytically by using an adiabatic equation of state, the Bernoulli equation, and mass continuity:
\begin{equation}\label{eq:adia}
    P_{x} = a\,\rho_{x}^{\gamma}
\end{equation}
\begin{equation}
    \frac{1}{2}v_{x}^2 + \frac{\gamma}{(\gamma-1)}\frac{P_{x}}{\rho_{x}}=b
\end{equation}
\begin{equation}
    \pi\,w^{2}\,\rho_{x}\,v_{x}=c,
\end{equation}
where $w$ is the width of the cavity, and $a,\,b,\,c$ are constants. The values of $b$ and $c$
can be expressed in terms of $a$:
\begin{equation}
  b=\frac{\gamma}{\gamma-1}a^{1/\gamma}\left( \rho_{0}v_{\star}^{2} \right)^{1-1/\gamma},
\end{equation}
\begin{equation}
  c=\frac{\gamma-1}{\gamma}\dot{E}a^{-1/\gamma}\left( \rho_{0}v_{\star}^2 \right)^{1/\gamma-1}.
\end{equation}

In \citep{Yoon:11}, the value of $a$ was a free parameter due to the complexity in jet geometry,
however the spherical shape of pulsar winds allows us to determine $a$ from pressure balance
at the standoff position:
\begin{equation}
    P_{\rm st,wind}=a\,\rho_{\rm st,wind}^{\gamma}=P_{\rm ram}=\rho_{0}\,v_{\star}^{2}.
\end{equation}
where $\rho_{\rm st,wind} = \dot{E}/ (4\pi R_{\rm st}^2 v_{\rm wind}^3)$ is the wind density
at the standoff position. As a result, $a$ can be expressed as
\begin{equation}\label{eq:aa}
    a=\frac{\rho_{0}v_{\star}^2}{\rho_{\rm st}^{\gamma}}
     = \left( \rho_{0}v_{\star}^{2}\right)^{1-\gamma}\,v_{\rm wind}^{2\gamma}.
\end{equation}

Using eqns.~(\ref{eq:adia})-(\ref{eq:aa}), $w_{1}$ can be obtained if we assume the pulsar is
moving supersonically ({\it i.e.} $P_{x} \ll \rho_{0}v_{\star}^{2}$):
\begin{equation}
    w \simeq \left(\frac{1}{2\pi^{2}}\right)^{1/4}\left( \frac{\gamma-1}{\gamma} \right)^{3/4}\dot{E}^{1/2}
    \,v_{\rm wind}^{-1/2}\left( \rho_{0}v_{\star}^{2} \right)^{1/2\gamma-1/2}\,P_{x}^{-1/2\gamma} = A\,P_{x}^{-1/2\gamma}.
\end{equation}

The pressure can be calculated by the jump conditions for an oblique shock [detailed derivation
is included in eqns.(13)-(18) in \citet{Yoon:11}], and the differential equation for the width
can be expressed as
\begin{equation}\label{eq:dwdx}
    \frac{dw}{dx}
    = \left( \frac{\gamma}{2} \right)^{-1/2}{{\mathcal M}_{0}}^{-1}
    \left[ \frac{(w/A)^{-2\gamma}}{P_{0}} -1 \right] 
    \left[
      \frac{\gamma+1}{P_{0}}\left(\frac{w}{A}\right)^{-2\gamma}+\gamma-1
    \right]^{-1/2},
\end{equation}
where $P_{0}$ is the ambient pressure and ${\mathcal M}_{0}=v_{\star}/\sqrt{\gamma\,P_{0}/\rho_{0}}$
is the Mach number of the pulsar relative to the ISM.

\begin{figure}
   \centering
   \includegraphics[width=\columnwidth]{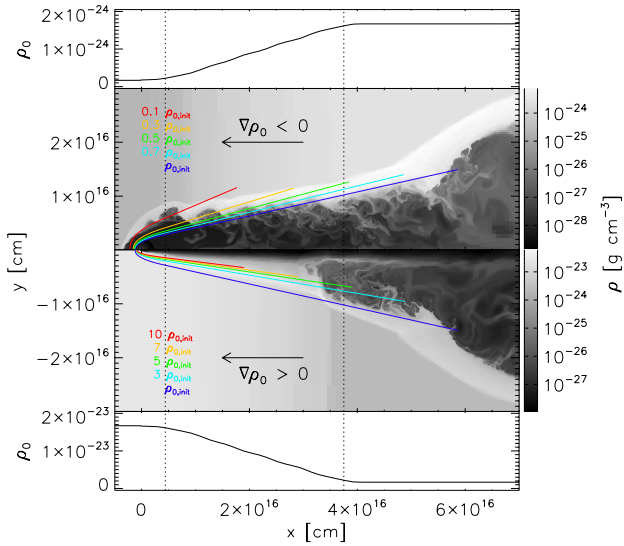}
   \caption{Density slice of the PWN head when a
     pulsar passes through either 10 fold increase (upper) or 1/10
     fold decrease (lower) in ambient density. The top and bottom
     plots show the smooth variation of the ambient density with
     a transition width of $5\times10^{17}\,{\rm cm}$
     (L10br\_600 \& H10br\_600 in Table~\ref{tab:param}). The
     transition locations are marked by vertical
     dashed lines.  The colored lines indicate analytic bow-shock
     solutions with different ambient densities.}
\label{fig:head}
\end{figure}

Integrating eq.~(\ref{eq:dwdx}) yields the analytic shape of the bow-shock and the neck from the
physical properties of the pulsar (ambient density, ambient pressure, spin-down loss energy rate,
wind velocity, and adiabatic index of the pulsar wind), and is consistent with our numerical
results (Figure~\ref{fig:head}).

This semi-analytic solution is helpful in understanding the morphological evolution of a bow-shock
head and a trailing neck when the pulsar moves through a medium in which the density changes
gradually.  If the ambient density {\em increases} gradually along the pulsar's trajectory, the ram
pressure also increases. Thus, the neck (analytic lines from eq.~(\ref{eq:dwdx}) are plotted in
Figure~\ref{fig:head}) becomes thinner. If the density of the ambient density {\em decreases}, the
bow-shock and the neck become wider.  We will apply this analysis to the Guitar Nebula's head in
\S\ref{subsubsec:head}.

\subsection{Pulsar wind evolution in density
gradients}\label{subsec:discontinuity}

When the pulsar encounters in increase in the ambient density, the width and opening angle of
the neck always decrease, constricting the flow.  No new bubble forms in this case, as expected.

\subsubsection{Density discontinuities}

When the density {\em decreases}, the width and opening angle of the neck increase.  We carried out
2-D hydrodynamic simulations to test if a bubble forms when the pulsar passes through a medium in
which the density changes sharply ({\it i.e.,} density discontinuity). We set the discontinuity to
be perpendicular to the pulsar's velocity.

In this case, the standoff distance $R_{0}$ increases [eq.~(\ref{eq:std})] rapidly and the bow-shock
becomes wider, producing a bubble at the discontinuity.  Figure~\ref{fig:evol} shows that the
formation and the evolution of a bubble proceeds in a similar fashion to the pulsar's initial
inflation of a bubble.

In the early stage after encountering the low-density medium, the expansion velocity of the bubble
is larger than the pulsar's propagation velocity, thus the pulsar travels inside the expansing
bubble. In this case (unlike the initial evolution of a pulsar in uniform medium), the isotropic
pulsar wind generates a {\em hemi-spherical} bubble, since the down-wind half of the pulsar wind is
impeded by the high-density medium. The analytic model of the expanding bubble can be applied to the
expanding forward-half of the bubble:
\begin{equation}\label{eq:bubble}
   R_{b}(t) = \eta^{1/5} \left( \frac{\dot{E}}{\rho_{0}} \right)^{1/5} t^{3/5},
\end{equation}
where $\eta$ is a constant that equals $125/154\pi$, which is derived by the assumption that
the gas is swept-up into a cold, thin shell \citep{Castor:75}.

The expansion velocity slows down with time while the pulsar's velocity is constant, and after
a time $t_{\rm break}$, the pulsar breaks out of the bubble, which can be calculated by the
bubble radius [eq.~(\ref{eq:bubble})] and the pulsar's velocity:
\begin{equation}
  t_{\rm break} = \eta^{1/2} \left( \frac{\dot{E}}{\rho_{0}} \right)^{1/2} v_{\star}^{-5/2}.
\end{equation}
at a bubble radius 
\begin{equation}
  R_{\rm break} = t_{\rm break}v_{\star}
  \label{eq:rbreak}
\end{equation}

Beyond this point, the pulsar's wind once again generates a bow-shock, a trailing conical neck,
and a terminal (hemi-spherical) bubble.

\begin{figure}
   \centering
   \includegraphics[width=\columnwidth]{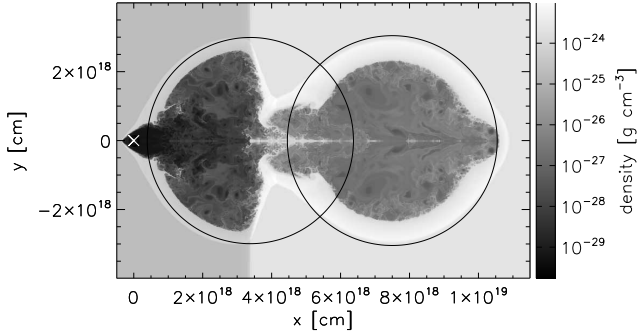}
   \caption{Density contour map when the pulsar penetrates the density
     discontinuity toward lower ambient density (model L10\_600).  The
     black circles are analytic models of the bubbles:
     the right one is produced by initial explosion
     and left one is produced when the pulsar encounters the density
     discontinuity. The white cross indicates the location of the
     pulsar.}
\label{fig:denlow}
\end{figure}

Figure \ref{fig:denlow} shows a well-developed double-bubble structure behind the bow-shock
head. The first bubble originated from the initial launch of the pulsar wind [and is therefore
not a result of the density structure of the ambient medium --- its position and size will depend
on the age and velocity of the pulsar according to eqs.~(\ref{eq:bubble}) and (\ref{eq:rbreak})];
the second bubble is generated when the pulsar encounters the discontinuity.

Note the small ``protrusion'' on the propagation axis on the right side of the initial bubble. This
is a 2-D numerical artifact that is caused by focusing the downstream flows along symmetry-axis
(i.e. $r$=0), impacting the bubble edge.  In the 3-D test run, this protrusion disappears. However,
given the small volume, the effects of this structure on the evolution of the bubble is negligible.

For comparison, the white cross indicates the location of the pulsar, and the black circles
represent the expanding bubbles in the analytic model [eq.~(\ref{eq:bubble})].

\begin{figure}
   \centering
   \includegraphics[width=\columnwidth]{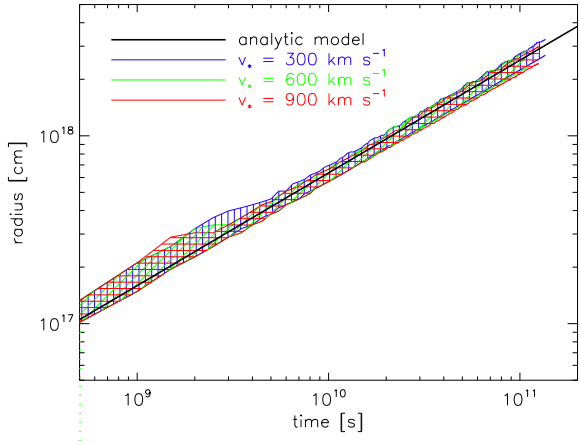}
   \caption{The time evolution of the expanding bubbles. The black
       solid line represents the analytic solution [eq.~(\ref{eq:bubble})], 
       and the colored areas represent the bubble
     shells between inner and outer radius of the bubble from the
     simulation results with $v_{\star}=300,\,600,\,900\,{\rm km\,s^{-1}}$.}
\label{fig:bubble_anal}
\end{figure}

We carried out simulations with pulsar proper velocities of 300, 600, and 900 $\rm km\,s^{-1}$.
Figure \ref{fig:bubble_anal} shows the evolution of the expanding bubbles with time for the
different velocity cases.  The solid line represents the analytic model in eq.~(\ref{eq:bubble}), and
each filled region represents the location of the shell of the bubble\footnote{Note that in this
work, we neglect cooling, broadening the shell at the surface of the bubble. If cooling is taken
into account, the shell thins, but the overall dynamic evolution is not affected \citep{Yoon:11}.}
For all cases, the evolving bubbles in the simulations are consistent with the analytic model.

If multiple bubbles are observed in a bow-shock PWN, we can constrain the parameters of the density
jump from the bubble's geometry.  In our simulations, we fixed the separation between the initial
position of the pulsar and the density discontinuity to be $l_{sep} = 4.12\times10^{18} \rm \,cm$.
This implies that if the pulsar moves faster, it reached the discontinuity in a shorter time,
thereby the size ratio of the second bubble to the first bubble becomes larger compared to the case
of slowly moving pulsars.  The ratio of the bubble size can be derived from the ratio in
eq.~(\ref{eq:bubble}),
\begin{eqnarray}\label{eq:bubratio}
    R_{b2}(t) / R_{b1}(t) &=& \left( \frac{\rho_{0,1}}{\rho_{0,2}} \right)^{1/5} \left( 1 -
    \frac{t_{d}}{t} \right)^{3/5} \nonumber \\
	  &=& \left( \frac{\rho_{0,1}}{\rho_{0,2}} \right)^{1/5} \left( 1 - \frac{l_{\rm sep}}{t\,v_{\star}}
	  \right)^{3/5},
\end{eqnarray} 
where $\rho_{0,1}, \rho_{0,2}$ are the ambient densities surrounding the first and second bubble,
respectively, and $t_{d}$ is time when the pulsar encounters the density continuity. The time
is related to the separation between the two bubbles by $l_{\rm sep} = t_{d}\times v_{\star}$.
This robust expression is useful for estimating the density changes in the medium if multiple
bubbles are observed.

\subsubsection{Secular changes in density}
\label{sec:secular}

We also simulated the case of a pulsar moving through a {\em smooth} variation of the ambient
density, with characteristic transition length $\Delta l$ over which the density changes, such that
\begin{equation}
  \frac{d\rho}{dx}\sim \frac{\rho}{\Delta l}
\end{equation}
In this case, the morphology of the bow-shock and the neck adjusts smoothly to the changing
density, as can be seen in Figure~\ref{fig:head}.

In this case, a bubble is {\em not} formed as long as $\Delta l \gg R_{\rm break}$ from
eq.~(\ref{eq:rbreak}).

\section{Discussion}

\subsection{Guitar Nebula}

\begin{figure}
    \centering
    \includegraphics[width=\columnwidth]{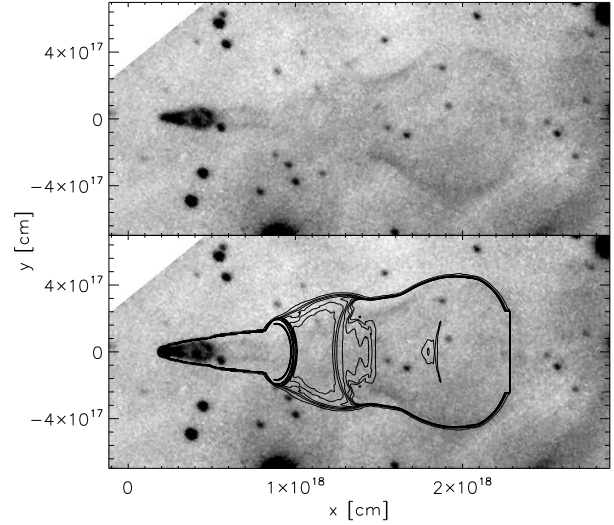}
    \caption{Guitar Nebula in H$_{\alpha}$, imaged with the 5m Hale
      Telescope at Palomar Observatory \citep{Chatterjee:02}. In
      the bottom panel, we overlaid the contour of
      H$_{\alpha}$ emission, which is calculated from the numerical
      results. See \S \ref{subsec:guitarbody}.}
\label{fig:guitarHa}
\end{figure}

The so-called Guitar Nebula is a well known pulsar wind nebula with a curious morphology.
The nebula, powered by the relativistic wind from the high-velocity pulsar, PSR B2224+65,
displays an H$_{\alpha}$ bow-shock and multiple shells/bubbles down-stream from the pulsar.

The characteristic spin-down age of the pulsar of 1.1 Myr \citep{Hobbs:04}, and the velocity
of $v_{\rm psr} \sim 1500\,{\rm km\,s^{-1}}$ imply that its birthplace is very far from its
current location, and thus that the two bubble-like structures {\em cannot} be due to the
initial inflation of the pulsar wind.  Given the observed parameters of the nebula, we therefore
tested whether the effects of a non-uniform density on the dynamical evolution of the nebula
can explain the observations.

We adopt the model parameters of the Guitar nebula from \citet[and references
therein]{Kerkwijk:08}: $\dot{E}=10^{33}\,\pwr$, $v_{\star}=1.5\times10^{8}\,{\rm cm\,s^{-1}}$. The
spin-down loss energy of the nebula is a few orders of magnitude lower than that of typical
H$_{\alpha}$ bow-shock PWNe \citep{Gotthelf:04}. However, the extremely fast motion
of the pulsar in the Guitar nebula and the likely low density of the environment enables PSR
B2224+65 to produce the visible bow-shock features despite of such low $\dot{E}$.

The pulsar is located slightly out of the galactic plane ($l=108.6^{\circ}, b=6.8^{\circ}$)
\citep{Cordes:93}, indicating that it is about 227 pc away from the plane if we assume the distance
of 1.8 kpc, estimated from the pulsar's dispersion measure of DM$=35.3\, {\rm pc\,cm^{-3}}$ and the
NE2001 electron density model \citep{Cordes:93,Cordes:02,Kerkwijk:08}.  Hence, we adjust the ambient
density and pressure in the Guitar Nebula model from Galactic mid-plane values, which drop to
$1.2\times 10^{-25}\,{\rm g\,cm^{-3}}$ and $1.7\times 10^{-12}\,{\rm dyne\,cm^{-2}}$, respectively
\citep{Cox:05}.  Note that since the ionized material along the line of sight likely biases the DM
upward, the distance may be over-estimated \citep{Chatterjee:04}.

\subsubsection{Bow-shock and neck}\label{subsubsec:head}

\citet{Chatterjee:04} observed morphological changes of the Guitar Nebula's head in H$_{\alpha}$
emission between two epochs spaced 7 years apart (1994, 2001).  Under the assumption of a
momentum-conserving bow-shock model \citep{Wilkin:96}, they modeled the shock front, concluding that
the ambient density at the nebula tip should decrease by $n_{\rm A}(2001)/n_{\rm A}(1994) \approx
0.7$, where $n_{\rm A}$ is the number density of the surrounding medium. However, this work is
purely analytic.

To test this conclusion, we carried out numerical simulations to understand how bow-shock and
trailing neck evolve if the pulsar moves through the medium with either gradual increase or
decrease in the ambient density.  As discussed in \S\ref{sec:secular}, Figure \ref{fig:head}
shows that if the pulsar moves through a medium with decreasing density ({\it i.e.}, $\nabla
\rho_{0}<0$), the standoff radius becomes larger, and the shape of bow-shock becomes rounder,
producing a wider neck.

On the other hand, if the pulsar moves through a medium with increasing density ({\it i.e.},
$\nabla \rho_{0}>0$), the stagnation point/standoff distance moves closer to the pulsar, and
the neck becomes thinner than in the initial density medium. The analytic lines in the figure
describe the change of in the shape of the shock fronts along the density variation in the medium.

\begin{figure}
    \centering
    \includegraphics[width=\columnwidth]{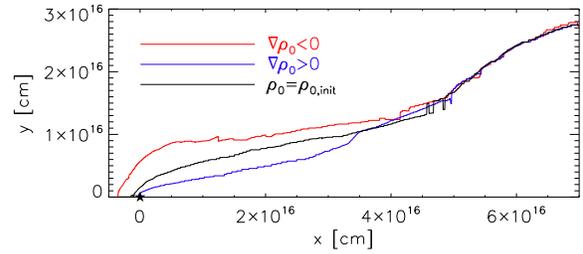}
    \caption{Comparison of the observed shape of the bow-shock head of
      the guitar nebula with simulation results. The red, blue and
      black represent the bow-shock and the trailing neck from the
      model of L10br\_600, H10br\_600, and Uniform\_600,
      respectively. The black star symbol indicates the location of
      the pulsar.}
\label{fig:head_comp}
\end{figure}

Figure~\ref{fig:head_comp} shows that the morphology of the bow-shock and neck differs between
the three cases: $\nabla \rho_{0}<0$, $\nabla \rho_{0}>0$, and constant $\rho_{0}$.  We found
that the rounder shape in the tip of the bow-shock head and the flattened neck in case of $\nabla
\rho_{0}<0$ is in good agreement with observed morphology of the Guitar head at 2001 [See Figure
2 in \citep{Chatterjee:04}], indicating that the pulsar likely moves decreased-density medium
during the epochs.

Under the assumption of constant spin-down loss rate $\dot{E}$, spatial velocity, and pulsar
wind velocity, we can obtain the density ratio in the medium by using eq.~(\ref{eq:std}), into
which we plug the ratio of standoff data, $\theta_{0, 2001}/\theta_{0,1994} \approx 1.25$,
where $\theta_{0}$ is a modeled standoff angle \citep{Chatterjee:04}.  The resultant density
ratio, $\rho_{0,2001}/\rho_{0,1994}$, is 0.64 which is consistent with \citet{Chatterjee:04}.

\subsubsection{Guitar Body}\label{subsec:guitarbody}

The origin of the bubbles behind the pulsar in the Guitar Nebula has been elusive because the
pulsar is far away from its birth place, the Cygnus OB3 association \citep{Tetzlaff:09}, so the
initial explosion cannot be responsible. However, as we discussed in \S~\ref{subsec:discontinuity},
an expanding bubble can be produced if the pulsar encounters a density discontinuity at which
the density sharply decrease.

We adopt the angular separation and angular size of the bubbles that were reported by
\citet{Kerkwijk:08}. We assumed a distance of $d=1.8\,{\rm kpc}$. The separation length from the
pulsar to the first and second bubbles (named in order of decreasing distance from the pulsar) can
then be calculated as $l_{\rm b1}=1.75\times 10^{18}\,{\rm cm}$ and $l_{\rm b2}=1.1\times
10^{18}\,{\rm cm}$, respectively.  Given the pulsar proper velocity,
$v_{\star}=1.5\times10^{8}\,{\rm cm\,s^{-1}}$, we expect that the first and second bubbles were born
370 and 233 years ago, respectively, while the pulsar passes through a non-uniform density medium.

Suppose that each bubble is produced when the pulsar passes through a density discontinuity. Then,
in order to reproduce multiple bubbles in the Guitar Nebula, we can calculate the change of ambient
density required to explain the bubbles.  Using the energy-driven bubble model
[eq.~(\ref{eq:bubble})], the ratio of the ambient density around the second bubble ($\rho_{2}$) and
first bubble ($\rho_{1}$) can be expressed as,
\begin{equation}
  \frac{\rho_{2}}{\rho_{1}}=\left( \frac{R_{\rm b1}}{R_{\rm b2}} \right)^5 \left( \frac{l_{\rm
        b2}}{l_{\rm b1}} \right)^3,
\end{equation}
where $R$ is the bubble radius. 

Given the parameters for the two bubbles in the guitar nebula \citep{Kerkwijk:08}, the ratio of the
bubble radius is $R_{\rm b1}/R_{\rm b2}=16''/9''\approx 1.78$, and the ratio of the separation is
$l_{\rm b1}/l_{\rm b2}=65''/41''\approx 1.58$.  This implies that the density around the second
bubble should be 4.5 times {\em higher} than that around the first bubble. Since a higher density
medium in the pulsar's path does not produce a bubble, the presence of the second (smaller) bubble
cannot be explained without another region of even higher density between the two media (i.e., a
thin wall) with density higher than $\rho_{2}$.

Such a complicated density structure would require a significant level of fine-tuning. Furthermore,
in this model, the H$_{\alpha}$ emission in the second bubble should be brighter than that in
the first bubble due to the enhanced ambient density. This is inconsistent with the observed
features, at which the H$_{\alpha}$ emission of the second bubble-like structure is a factor
of 5 dimmer than that of the first bubble \citep{Chatterjee:02}.

Based on the intuition gained from the simulations presented above, we propose the following
scenario for producing the guitar shape of the nebula as the result of the pulsar traversing a
region of low and decreasing density before re-entering a higher density region: (1) The first
(larger) bubble is produced when the pulsar passes through a low density discontinuity. (2)
The second bubble is produced as the pulsar moves along the decreasing density gradient. (3)
The neck requires passage into a region of higher density.

\begin{figure*}
    \begin{center}$ 
        \begin{array}{cc} 
              \includegraphics[width=0.55\textwidth]{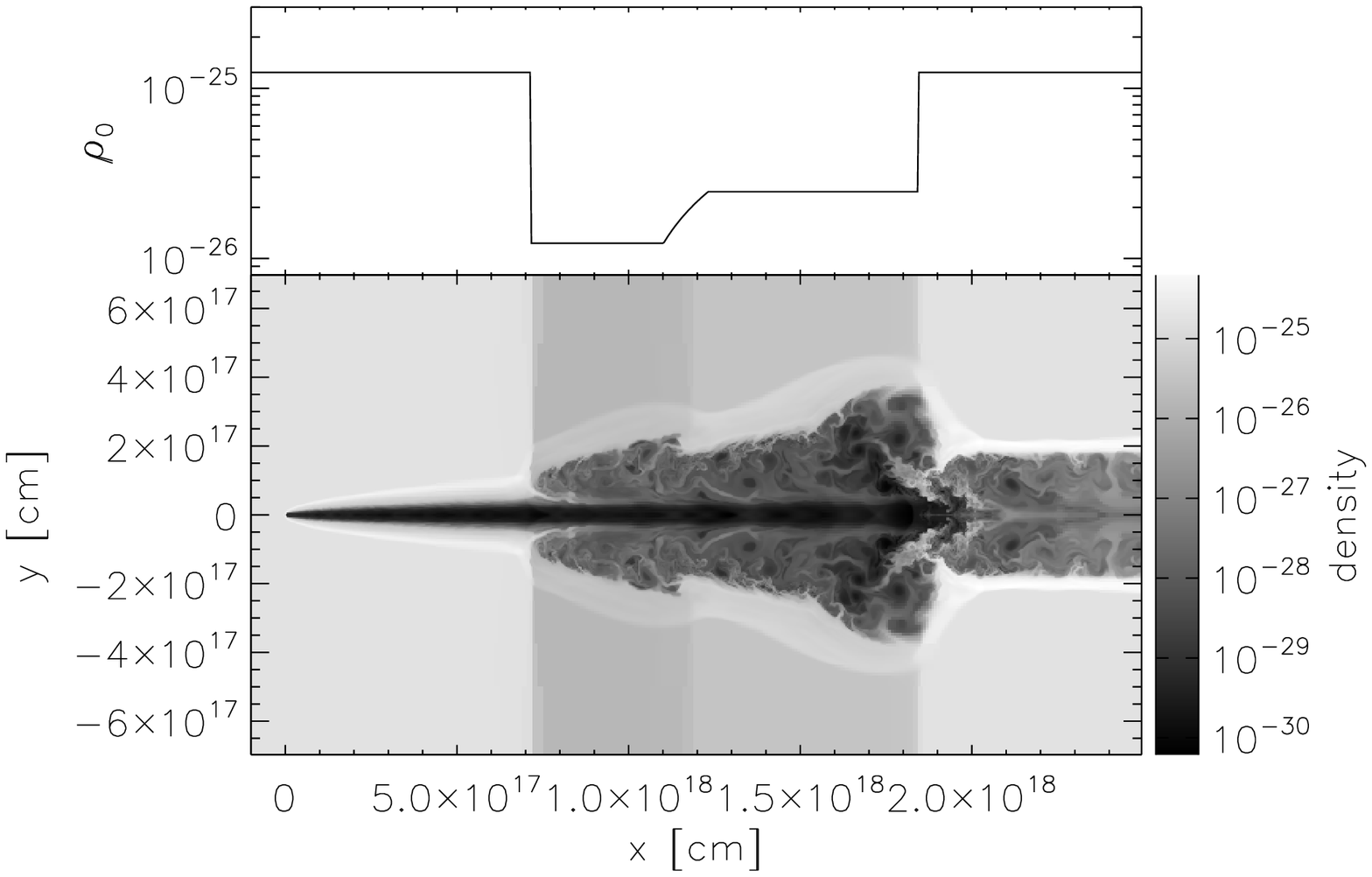} &  
              \includegraphics[width=0.45\textwidth]{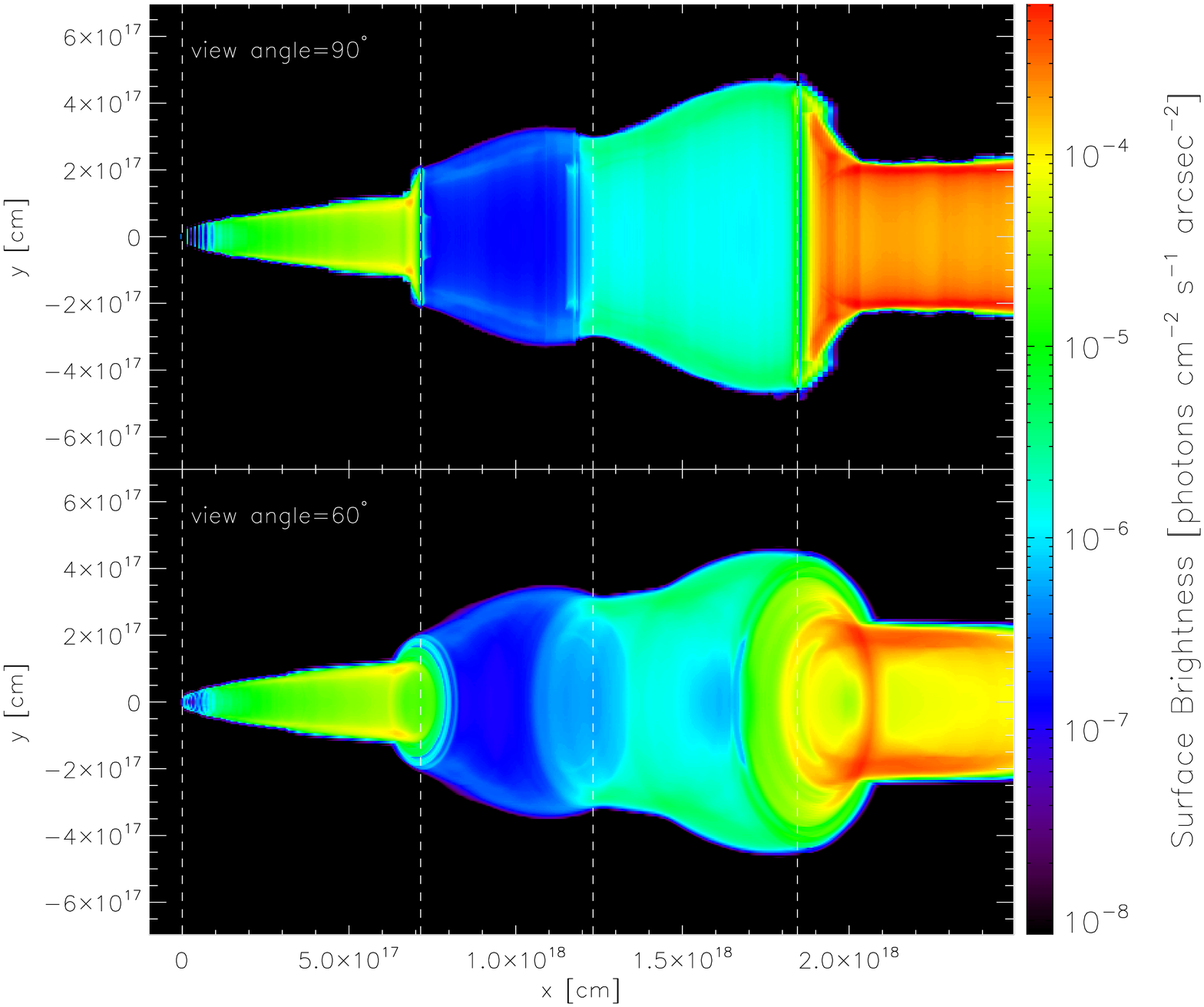}    
           \end{array}$ 
     \end{center} 
     \caption{Left panel:Density contour map for the Guitar Nebula model (model Guitar\_LOS90). The variation of ambient density 
             was described in upper plot. Right panel: Projected H$_{\alpha}$ emissions in logarithmic scale with the viewing angle 
             of 90$^{\circ}$(upper) from the model Guitar\_LOS90 and 60$^{\circ}$(lower) from the model Guitar\_LOS60.
             The left most vertical dashed line indicates the pulsar's position and other vertical lines indicate the locations of density change in the medium.}
     \label{fig:guitar} 
\end{figure*}

We carried out two sets of 2-D hydrodynamic simulation to test this scenario to reproduce the aspect
ratio of the guitar for two different viewing angles, (model Guitar\_LOS90 seen edge-on, and
Guitar\_LOS60 seen at 60$^{\circ}$). The ambient density variation is shown in the upper plot of
Figure~\ref{fig:guitar}.  The density contour map in Figure~\ref{fig:guitar} shows that this model
reproduces the shape of the guitar body and the narrow neck with reasonable precision.

In order to compare this result to observed H$_\alpha$ emission, under the assumption of collisional
ionization equilibrium, we calculated the emission from the numerical result.  Note that we
generated 3-D data based on the 2-D axi-symmetric results to obtain the projected H$_{\alpha}$
emission.

The ionization balance in the shocked gas was calculated by the MAPPING III code
\citep{Sutherland:93}. The temperature of the shocked shell is above $10^{6}$ K, thus the
gas is nearly fully ionized.  The right panel of Figure~\ref{fig:guitar} shows the surface
brightness in H$_{\alpha}$ for the model. The surface brightness around the rim of the bubble
is about $10^{-5}~{\rm photons\, cm^{-2}\, s^{-1}\, arcsec^{-2}}$, which is consistent with
the H$_{\alpha}$ observation \citep{Kerkwijk:08}.

The H$_\alpha$ emission of the inflated structure between the bubble and the pulsar is a factor
of 7 dimmer than that of the rim of the bubble due to the low density in the medium, which is
consistent with the observation.

Note that the bright emission far down-stream beyond the bubbles (on the fourth dashed line on the
right-hand-side of the image) does not have a corresponding feature in the observed map. However,
given that we cannot constrain the ratio of the density in this region of the flow other than
saying that it must be higher than in region (b), since no observed features have been identified
in this region of the flow, the simulation should not be expected to reproduce the results.

The bubble is flattened toward the discontinuity because the high density suppresses the expansion
of the bubble in this region. However, its shape can project into a circle if the viewing angle,
$\theta_{\rm LOS}$, between the pulsar velocity vector and the line of sight (LOS) varies (see right
panel of Figure~\ref{fig:guitar}).  Although $\theta_{\rm LOS}\sim 90^{\circ}$ is preferred because
otherwise the three-dimensional velocity of the pulsar can be extremely high, the viewing angle is
not well constrained between $30^{\circ}-60^{\circ}$ \citep{Chatterjee:02,Chatterjee:04}.  In the
model Guitar\_LOS60, we set the pulsar's velocity of $v_{\star}= 1.5\times 10^{8} /\sin{60^{\circ}}
\,{\rm cm\,s^{-1}}$, and the location of the density variation from the pulsar to be farther away by
$1/\sin{60^{\circ}}$.  We rotated the data with the viewing angle of $\theta_{\rm LOS}=60^{\circ}$,
and the projected H$_{\alpha}$ surface brightness shows the round shape of the bubble which provides
a better match to the observed shape. We overlaid the contour of the simulated surface brightness on
the observed H$_{\alpha}$ image, which is in good agreement with the observed shape (bottom panel of
Figure~\ref{fig:guitarHa})

The ambient density around the bubble was set using eq.~(\ref{eq:bubble}) 
and the observed properties of the pulsar wind and bubble:
\begin{equation}\label{eq:den0}
    \rho_{0} = \frac{\eta\, \dot{E}\,t^{3}}{R_{b}(t)^{5}}
        =\frac{\eta\,\dot{E}\,\Theta_{l,\perp}^{3}}
              {\Theta_{b}^{5} \, \mu_{\star,\perp}^{3} d^{5}}\simeq 2.5\times10^{-26}\,{\rm g\,cm^{-3}},
\end{equation}
where, $\Theta_{l,\perp}$ is the angular separation between the bubble center and the pulsar,
and $\Theta_{b}$ is the angular radius of the bubble, and $\mu_{\star,\perp}$ is the transverse
angular velocity of the pulsar. The angular velocity is estimated to $\mu_{\star,\perp}\simeq
182\pm3\, {\rm mas\,yr^{-1}}$ \citep{Harrison:93}. The resulting hydrogen number density can
be calculated from the density above to $n_{\rm H}\simeq 0.01$, which is consistent with the
nominal ambient density of the nebula, reported by \citet{Chatterjee:02}.

Since the ISM exists in multiple phases depending both in temperature and chemical composition
\citep{Draine:11}, accurately representing the density structure around the pulsar would require an
unreasonable degree of fine-tuning and carry a large amount of uncertainty. The chosen density
structure in the simulations (plane parallel with simple uniform regions) is clearly an
over-simplification. However, given the multi-phase nature of the ISM it is reasonable that the
pulsar will transverse density inhomogeneities of scale similar to the size of the guitar (roughly
of order a parsec). For example, a cluster of stellar winds could generate a local underdensity in
the medium that would correspond to the simulated density structure to lowest order.

Given the uncertainties in the observed physical quantities, eq.~(\ref{eq:den0}), we can estimate
the uncertainty of the density to be $\sigma_{\rm rho} \sim 130\%\,\rho_{0}$.  The dominant factor
determining the uncertainty in the density is the distance, $d$: the distance is obtained from the
pulsar dispersion measure considering a model of the Galactic electron density \citep{Taylor:93},
and it is constrained to within roughly 25\% uncertainty to be $d=2\pm0.5 \,{\rm
kpc}$\citep{Chatterjee:02}.  The structure of the density discontinuity is unconstrained on scales
significantly later than the guitar body size.

In the simulation, we set three different ambient densities along the pulsar's path to establish the
observed structures: thin neck, middle bubble-like structure, and the right-most expanding bubble
(see Figure~\ref{fig:guitar}). The corresponding density ratio of each region is
$(\rho_{0,1},\,\rho_{0,2},\,\rho_{0,3})=(5,\,0.5,\,1)\,\rho_{0,3}$, where the subscripts represent
each region from left to right and $\rho_{0,3} = 2.5\times 10^{-26} \,{\rm g\,cm^{-3}}$
[eq.~(\ref{eq:den0})].  Considering the error in the distance, the uncertainties of the density
ratio are estimated to be $\sigma_{\rho_{0,1}/\rho_{0,3}} \sim 8\%\,\rho_{0,1}/\rho_{0,3}$ and
$\sigma_{\rho_{0,2}/\rho_{0,3}} \sim 70\%\,\rho_{0,2}/\rho_{0,3}$, respectively.

\subsection{Asymmetric Shape of Bow-shock}
\label{subsec:asym}

\begin{figure*}
    \centering
    \includegraphics[width=\textwidth]{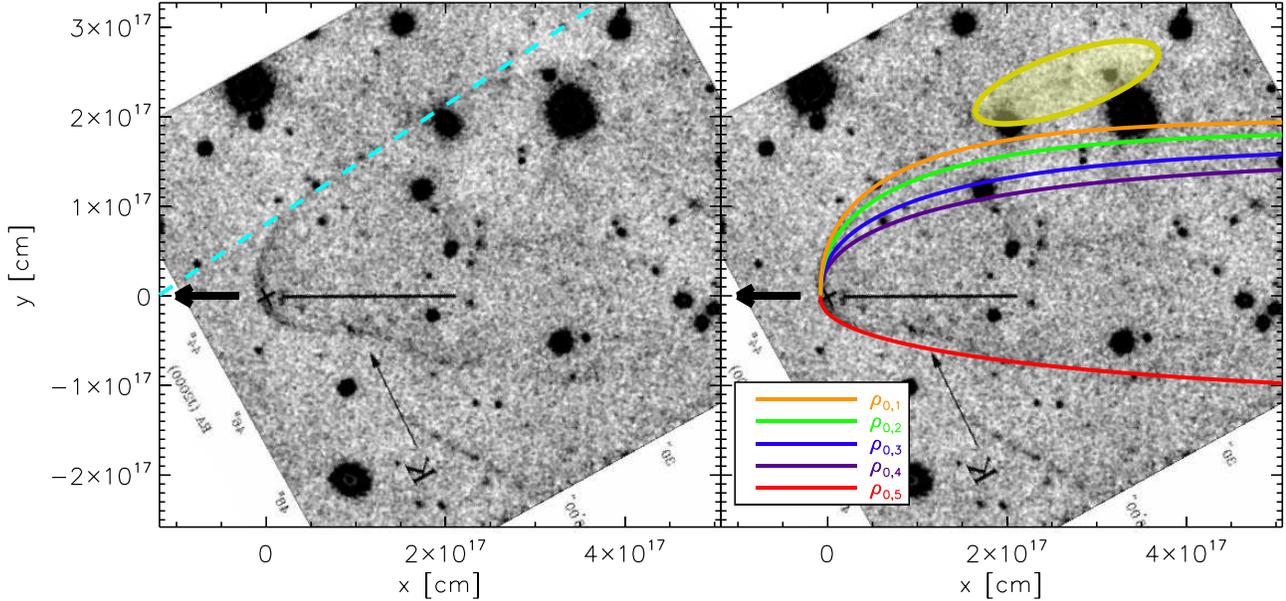}
    \caption{H$_{\alpha}$ image of PSR J2124-3358
      \citep{Gaensler:02}. In the left panel, the
      cyan dashed line represents the maximum opening angle of the bow-shock, 
      which is $\sim 30^{\circ}$. In the right
      panel, we overlaid the analytic bow-shock lines with several
      ambient densities on the image:
      $\rho_{0,1}=2.5\times 10^{-25}\,{\rm
        cm},\,\rho_{0,2}=4.2\times10^{-25}\,{\rm cm},
      \,\rho_{0,3}=9.2\times10^{-25}\,{\rm
        cm},\,\rho_{0,4}=1.7\times10^{-24}\,{\rm cm},
      \,\rho_{0,5}=1\times10^{-23}\,{\rm cm}$.
      The yellow circle represents the area where the analytic
      line deviates from the observed bow-shock.}
    \label{fig:gaensler}
\end{figure*}

The H$_\alpha$-emitting bow-shock nebula powered by the nearby millisecond pulsar J2124-3358 has a
highly asymmetric shape in its head around the pulsar's velocity vector (Figure~\ref{fig:gaensler}).
Assuming that the velocity vector of the pulsar is perpendicular to the line of sight, the standoff
angle is estimated to $2''.6$, implying that the standoff distance is $R_{0} \approx
3.4\times10^{-3}\,{\rm pc}$ with the distance of $\approx 270\, {pc}$ \citep{Cordes:02}. The upper
bow-shock is about 3.5 times wider than the lower bow-shock, and the upper one has a broader shape.
The origin of the asymmetry has been elusive and cannot be explained by the interaction of an
isotropic pulsar wind with a homogeneous ambient medium.  More interestingly, a kink-like shape,
which is a factor of ~2 brighter than the nebula average, appears in the lower bow-shock (denoted by
``K'' in Figure~\ref{fig:gaensler}).

One possible reason for the asymmetry in the bow-shock is that the density in the region to the
North of the pulsar is lower than in the South, with the pulsar currently crossing a density
discontinuity that is inclined with respect to the pulsar’s velocity vector (different from
the geometry discussed so far).  If we assume that the asymmetry in PSR J2124-3358 is caused by
the differences in ambient density between the Northern and the Southern side, we can estimate the
relative densities from by-eye fits of analytic bow-shock models for different densities; five such
models are overlaid on the observed bow-shock (right panel of Figure~\ref{fig:gaensler}). For the
analytic lines, we adopted the pulsar's spin-down loss energy of $\dot{E}=4.3\times10^{33}\,\pwr$
\citep{Toscano:99} and the projected space velocity of $60\,{\rm km\, s^{-1}}$ \citep{Gaensler:02}.

This suggest that the ambient medium of the upper region may be approximately 40 times less dense
than that of the lower region, unless there is bulk motion of the ambient gas to decrease the angle
of the lower bow-shock.  The upper half of the bow-shock deviates from a simple analytic model
farther back along the flow (denoted by a yellow circle in the figure).  The analytic lines converge
to become parallel to the pulsar's velocity vector after $x=2\times10^{17}\,\rm cm$, because the
momentum flux of the ISM pushes the bow-shock backward due to the pulsar's proper motion.  However,
the observed bow-shock has an opening angle of $30^{\circ}$ (cyan dashed line in
Figure~\ref{fig:gaensler}).  This deviation may be caused by an overestimate in the distance to the
pulsar and/or the uncertainty in the viewing angle of the bow-shock, i.e., the angle between the
velocity vector and the line of sight: if the distance is shorter than estimated, the bow-shock of
downstream likely retains a larger opening angle in two reasons. First, in the case of a shorter
distance (i.e. smaller physical size of the observed structure), the bow-shock may not reach the
region where it becomes parallel to the pulsar's proper motion. Second, the pulsar's velocity should
be lower than the estimated one, implying that the opening angle of the bow-shock is likely wider
than what we can see from the analytic lines in the figure.

Alternatively, this mismatch may be produced by the formation of flow instabilities as the pulsar
passes the density gradient. This will be discussed in detail below.

\citet{Gaensler:02} argued that the asymmetric bow-shock structure is caused by the combination of
three components: a density gradient in the ISM, a bulk flow in the ambient medium, and an
anisotropy in the pulsar wind. However, in their semi-analytic models for the structure, an
inclination angle between the pulsar's velocity vector and the density discontinuity cannot be taken
into consideration due to the complexity. In the models, the density gradient was either
perpendicular or parallel.

To study the effect of the inclination angle $\varphi$ of the density discontinuity relative to the
proper motion of the pulsar, we carried out a 3-D hydrodynamic simulation of a bow-shock PWN in
which the pulsar moves into a lower-density medium such that its velocity vector is inclined with
respect to the transition layer at an angle of $\varphi = 5^\circ$.

This angle is chosen somewhat arbitrarily, but is set to be small enough to reproduce bow-shock
structures that are distinct between the shock generated at the low ambient density the pulsar is
propagating into and the shock generated at the initial, higher ambient density in the medium the
pulsar is initially traveling through.  If the angle is larger than the opening angle of the
bow-shock in the low-density medium (cyan dashed line in the right panel of
Figure~\ref{fig:gaensler}), the shape of the bow-shock will become more symmetric.  The rather
complicated shape of bow-shock in the PSR J2124-3358 (e.g., the presence of two kink-like
structures; see Figure~\ref{fig:gaensler}) requires a small inclination angle. The generation of the
complicated structures will be discussed in detail below.

Unlike the semi-analytic model in \citet{Gaensler:02}, our model focuses only on the variation
of the ambient density along the pulsar's passage without taking any anisotropy in the pulsar
wind into consideration.

An accurate to-scale 3-D simulation of the J2124-3358 bow-shock would require an unreasonable
amount of computing time for the bow-shock to establish, given the small Courant time step
imposed by the pulsar wind velocity. For computational expedience, we thus approximated the
system using a non-relativistic 3-D model similar to the other models used in this paper, the
parameters of which are given in Table~\ref{tab:param} (L10\_1000\_5deg). The primary aim of
the simulations is to explore whether the asymmetry of pulsar wind bow-shocks can be produced
by simple inhomogeneities in the environment.

In our numerical model, the ambient density decreases to 1/10 of the initial density $\rho_{0}$
across a transition zone, with a smooth density transition over a length scale $\Delta l_{\rm
tr}$, such that the pulsar encounters a projected transition length scale
\begin{equation}
  \Delta l_{\rm tr,x}=\Delta l_{\rm
    tr}/\cos({5^{\circ}})=5\times10^{17} \rm cm
\end{equation}
in the direction of motion, i.e., the x-direction, which is 100 times larger than the standoff
distance of the pulsar wind.

The transition length is crucial for determining how rapidly the bow-shock expands while the density
is changing. If the transition length is small compared to standoff distance (i.e., $\Delta R_{0}/
\Delta l_{\rm tr,x} \gg 1$), as discussed in \S\ref{subsec:discontinuity}, the tip of the bow-shock
head inflates in a short time, producing an expanding bubble-like structure behind the pulsar.

However, if the transition length is sufficiently large, the evolution of the bow-shock head and
trail is smooth without bubble formation because the radius of curvature of the bow-shock head
increases and the trail becomes flat (see \S \ref{subsubsec:head}).

\begin{figure*}
   \centering
   \includegraphics[width=\textwidth]{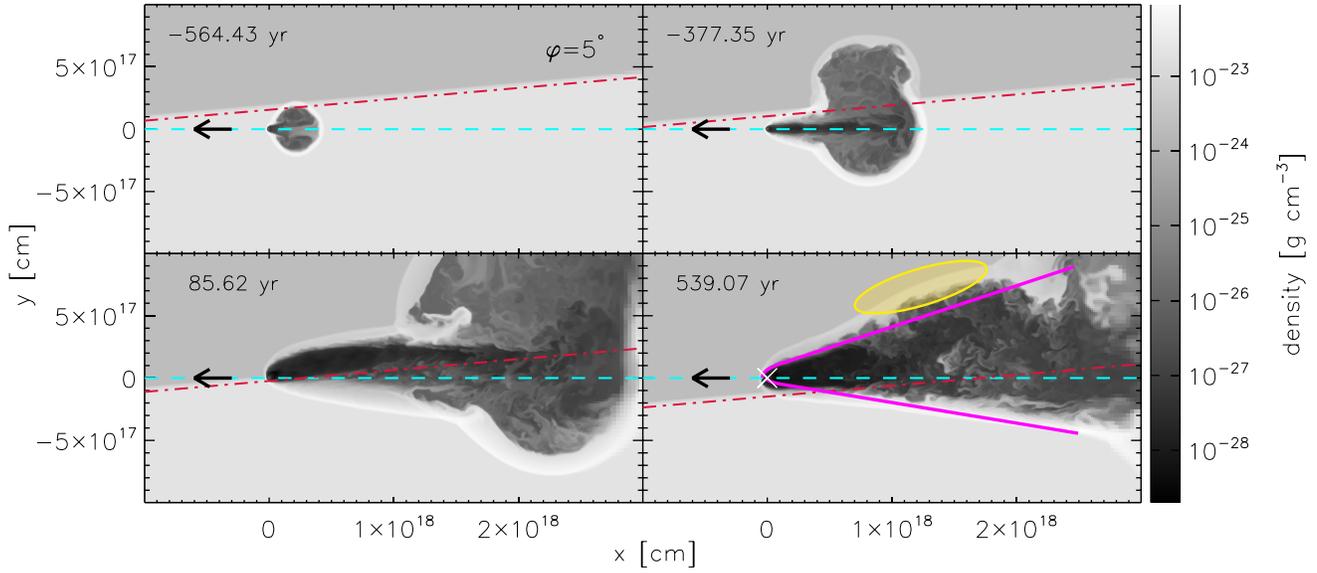}
   \caption{Density slices of the temporal evolution as
       the pulsar traverses a decrease in the density of the external
       medium which has an inclination angle of $5^{\circ}$ with
       respect to the pulsar's velocity vector (black arrow). The
     ambient density decreases to 1/10$\rho_{0}$ with
     a transition length of
     $\Delta l_{\rm tr,x}=100 R_{0}$, where $R_{0}$ is
     the standoff distance of the pulsar wind [see
     eq.~(\ref{eq:std})].  The bottom left panel shows the density map
     when the pulsar is within the transition layer. In the bottom
     right panel, the pulsar passed the layer, and the bow-shock has
     re-shaped.  The white cross indicates the location of the pulsar,
     and the cyan dashed line and red dot dashed line represent the
     pulsar's trajectory and the surface of density
     change, respectively.  The magenta lines represent the analytic
     bow-shock [eq.~(\ref{eq:dwdx})], in which the upper
     line is calculated for a decreased ambient
     density of $1/10\rho_{0}$ and the
     bottom line is calculated for the initial ambient
     density $\rho_{0}$.  The yellow area marks the region where the
     bow-shock deviates from the analytic
     solution.}
\label{fig:inc}
\end{figure*}

Figure~\ref{fig:inc} shows the temporal evolution of the density while the pulsar moves into the low
density medium.  The transition area is marked with a red dot-dashed line, and the white cross
represents the pulsar (bottom right panel).  The magenta lines are the analytic solutions for the
bow-shock in uniform medium with ambient density $\rho_{0}$ (lower line) and with 1/10 $\rho_{0}$
(upper line), respectively.

At early stages of the simulation, the bow-shock is symmetric. However, as expected, it becomes
asymmetric after the pulsar moves into the low density medium; the width of the neck $w$, and the
asymptotic bow-shock angle, $dw/dx$, are inversely-proportional to the ambient density.  We ignore
the evolution of the initial bubble, since it is unrelated to the crossing of the density
discontinuity (it is formed as the pulsar turns on at the start of the simulation, as discussed
above), and only focus on the properties of the bow-shock.  We note that in the yellow area, the
bow-shock shows a deformation that deviates somewhat from the analytic line.  This is because
although the transition length $\Delta l_{\rm tr,x}$ is long enough to avoid the generation of an
expanding bubble, the sudden change of stand-off distance shortly makes the shock unstable while the
pulsar is passing the transition zone.  It is plausible that the kink-like structure observed in the
bow-shock of J2124-3358 is produced by a similar instability at the intersection between the
bow-shock and the transition layer.

\begin{figure*}
    \centering
    \includegraphics[width=\textwidth]{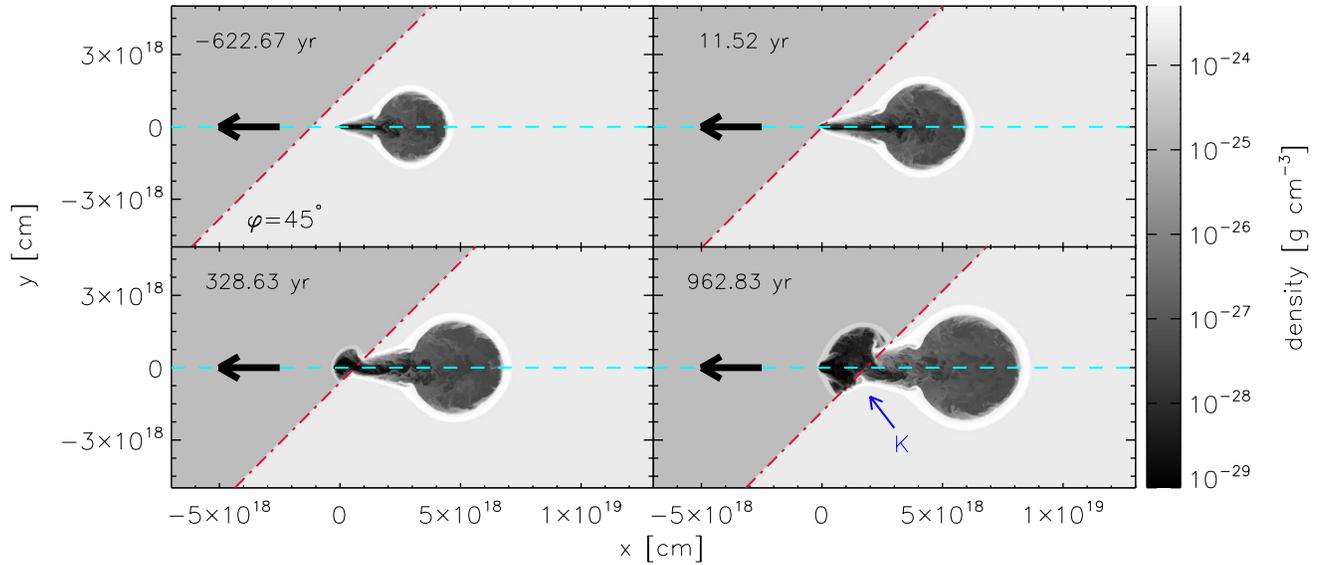}
    \caption{Temporal evolution of the density maps as
      the pulsar traverses a density decreased medium,
      which has the inclination angle of $45^{\circ}$ to the velocity
      vector.  The transition length is $\Delta l_{\rm tr,x}=0$. The
      rest configuration is same with the model in
      Figure~\ref{fig:inc}.}
    \label{fig:inc45}
\end{figure*}

For comparison, we also performed a simulation with the larger inclination angle of $\varphi =
45^{\circ}$ and a transition length of $\Delta l_{\rm tr,x}=0.$ Figure~\ref{fig:inc45} shows the
time evolution of the density map in this case. As we discussed above, the zero transition layer
(or, $\Delta l_{\rm tr,x} \ll R_{0}$) results in the production of expanding bubble rather than the
smooth change of bow-shock morphology. In this case, the asymmetric feature appears only when the
pulsar is in the bubble: After the pulsar breaks out of the bubble, it recedes from the transition
layer, producing a symmetric bow-shock.  However, the round shape of the bubble is not consistent
with the overall shock morphology of J2124-3358, although a kink-like structure appears at the
intersection between the lower bow-shock and the transition layer (denoted by ``K'' in the figure).

\subsection{Validation of the non-relativistic approximation}\label{subsec:relativistic}

Pulsar winds are magnetized and relativistic outflows, indicating that relativistic
magnetohydrodynamic (MHD) treatment is required in studying the {\em internal} flow structure and
the emission properties of bow-shock PWNe \citep{Bernstein:09}. However, even for the fastest
known pulsar, the pulsar velocity $v_{\star}$ is of 2 orders of magnitude less than the speed
of light, and the expansion velocity of the bubble and neck are correspondingly smaller, which
is manifestly non-relativistic.

We carried out some subsets of runs with the FLASH special relativistic hydrodynamic (RHD) solver to
compare with our results with non-relativistic hydrodynamic simulations.  In these test runs, we
varied the velocity of the pulsar wind ($v_{\rm wind}=0.33c,\,0.6c,\,0.9c,\,0.99c$) for both
non-relativistic and relativistic runs, and otherwise we used identical parameters.

\begin{figure}
   \centering
   \includegraphics[width=\columnwidth]{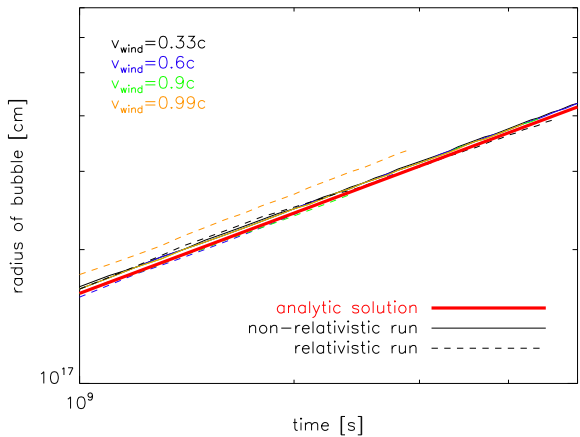}
   \caption{Results from the validation simulations of pulsar wind
     simulations. Shown is the time evolution of the radius of the
     expanding pulsar wind driven bubbles. The red solid line
     represents the analytic bubble solution [eq.~(\ref{eq:bubble})],
     and other colored lines represent the inner radius of the bubbles
     from the simulation results with
     $v_{\rm wind}=0.33c,\,0.6c,\,0.9c,\,0.99c$. The solid lines are
     non-relativistic hydrodynamic results, and the dashed lines are
     relativistic hydrodynamic results, showing that the scaling
     solution is in excellent agreement with the
     simulations and that relativistic corrections to the solutions
     are moderate.}
\label{fig:relbub}
\end{figure}

Figure \ref{fig:relbub} shows the evolution of pulsar-wind-inflated bubbles. For various pulsar wind
velocities, the trends of the bubble expansion are consistent with the analytic solution (red solid
line) from eq.~(\ref{eq:bubble}).  The error is about 8\% for the extreme case ($v_{\rm wind}=0.99c$),
and is less then 2\% for rest of the cases, validating our assumption that relativistic corrections
are moderate.

A fully relativistic MHD treatment is beyond the scope in this study. As we discussed, the
relativistic effect is negligible in the evolution of bubbles. As argues by \citet{Bucciantini:05},
the variation of a pulsar wind magnetization does not change the properties of the {\em external}
shock significantly, apart perhaps from the very head of the nebula.  Future simulations in
full MHD should verify this assumption.

\section{Conclusion}

In order to study the effect of inhomogeneous environments on the structure of bow-shock PWNe,
we performed a series of 2-D and 3-D hydrodynamic simulations.

We confirm the expectation that if a pulsar passes through a decreasing density gradient in the
medium, the standoff distance and radius of curvature of the bow-shock increase. As a result,
the overall shape of the trailing neck becomes more flattened during the transition. Conversely,
if the density in the medium is increasing, the radius of curvature and standoff distance of
the bow-shock decrease.  We derived analytic formula for the shape of the bow-shock head and
the trailing neck as a function of ambient density and pulsar wind parameters.

We applied this analysis to the head and bow-shock of the Guitar Nebula, and suggest that the
pulsar in the nebula is currently moving through a density-decrease.

If a pulsar encounters a density discontinuity at which the density decreases, the standoff distance
increases rapidly, producing a hemispherical bubble bounded by the high-density medium. Eventually,
the pulsar will break out of the bubble at radius $R_{\rm break}$, generating a characteristic
neck-bubble structure. We showed that the condition for bubble-formation is that the transition
length $\Delta l$ is smaller than the break-out radius $R_{\rm break}$.

We can reproduce the observed shape of the Guitar Nebula from a series of density changes in the
medium which suggests that the pulsar traversed a lower density region that gave rise to the body of
the guitar nebula.  Projection effects of the hemispherical bubble can explain the round appearance
of the body of the guitar nebula.

We showed that when the density discontinuity is inclined to the pulsar's velocity vector, it
generates an asymmetry in the shape of the bow-shock head, potentially explaining the observed
asymmetry in the bow-shock of PSR J2124-3358.

\section*{ACKNOWLEDGMENTS}
We would like to thank Bryan Gaensler, Enrico Ramirez-Ruiz, and Brian Morsony for helpful 
discussions. SH and DY acknowledge support form NSF grant AST 0908690.

\label{lastpage}

\end{document}